\documentclass[11pt,twoside]{article}
\usepackage{asp2004}
\usepackage{epsf}
\usepackage{psfig}
\usepackage{lscape}
\usepackage{graphics}
\begin{document}
\title{Discovery of The New WZ Sge Star, SDSS J080434.20+510349.2}
\author{E. Pavlenko$^{1,2}$, S. Shugarov$^{3}$, N. Katysheva$^{3}$, D. Nogami$^{4}$, K. Nakajima$^{5}$, H. Maehara$^{6}$, M. Andreev$^{7}$, V. Shimansky$^{8}$, A. Zubareva$^{8}$, Ju. Babina$^{2}$, N. Borisov$^{9}$, A. Golovin$^{10}$, A. Baklanov $^{1}$, D. Baklanova $^{1}$, K. Berezovsky$^{11}$, P. Kroll$^{12}$} 
\affil{$^{1}$ Crimean astrophysical observatory, Nauchny, Ukraine;
 email: guslik2000@mail.ru\\ 
$^{2}$ Tavrida National University, Simferopol', Ukraine\\ 
$^{3}$ Sternberg Astronimical Institute, Moscow, Russia\\
$^{4}$ Hida Observatory, Kyoto University, Kamitakara, Japan\\
$^{5}$ VSOLJ, Kumano, Japan\\
$^{6}$ VSOLJ, Saitama, Japan\\
$^{7}$ Terskol Branch of the RAS Institute of Astronomy, Russia\\
$^{8}$ Kazan State University, Kazan, Russia\\
$^{9}$ Special Astrophysical Observatory RAS, Nizhny Arkhyz, Russia\\
$^{10}$ Kyiv National Taras Shevchenko University, Kyiv, Ukraine\\
$^{11}$ Small Akademy of Science "Iskatel", Simferopol', Ukraine\\
$^{12}$ Sonneberg Observatory, Sonneberg, Germany\\}
\begin{abstract}
We present the results of the SDSS J080434.20+510349.2 photometry in its low state and during the outburst and spectroscopy during the outburst. We found such pecularities as the long-term outburst with amplitude probably not less than $6^{m}$, the rarity of the outbursts, eleven  rebrightenings,  short (0.059713(7) d) superhump period. We conclude this star to belong to the WZ Sge-type subclass of cataclysmic variables. The spectrum  shows both emission and absorption lines of H and He superimposed on the blue continuum. We also found 8 -- 9 min. brightness variations during the end of the superoutburst plateau that could be connected with the white dwarf pulsations in binary. 
\end{abstract}
\section{Introduction}
The SDSS J080434.20+510349.2 (hereafter SDSS0804) as a cataclysmic variable with underlying white dwarf was classified by Szkody et al. (2006) during its low ($\sim$ $18^{m}$) brightness state. The 0.0592(4) d orbital period  was found spectroscopically. The photometry yield the two-humped profile over the orbital period and  its amplitude increasing with brightness.
\section{Observations}
The photometry of the SDSS0804 has been carried out with different telescopes: 2.6-m Shajn telescope (ZTSh) and Cassegrain 38-cm telescope of CrAO, 70-cm, 60-cm and 50-cm telescopes of SAI, 60-cm telescope of Terscol, 25-cm telescopes of VSOLJ. Spectroscopy was done with 6-m BTA telescope of SAO. We present observations carried out in December, 2005 -- September, 2006. The brightness was measured relatively the comparison star TYC2 (3141.1011.1).
\section{The light curve of the outburst}
We started the observations of SDSS0804  when it was faint ($\sim 17^{m}.5$). On March, 4, 2006, we first have caught it in the outburst as the star of $\sim 13^{m}$ showing the  superhumps with period 0.06 d and amplitude $0^{m}.2$. Following observations shown the rapid faintness up $15^{m}.5$ in a five days, so we concluded  that we observed the star at the end of the superoutburst.
Taking into account that during the plateau of superoutburst the SU UMa stars typically fade on $\sim 1^{m} - 2^{m}$, one could suggest that in a top brightness SDSS0804 was $11^{m} - 12^{m}$. Hence the amplitude of the outburst was more or equal to $6^{m}$, that is typical for the WZ Sge stars.

Soon after discovery of the outburst we observed the  rapid brightness decline and series of the eleven rebrightenings. The mean amplitude of the separate rebrightening was $\sim 1^{m}.3$ and duration  of 2.5 d. After the end of the rebrightenings the era of slow decay with different rate began. The outburst light curve is shown in Fig.1. Note that only a few of WZ Sge stars also displayed the series of rebrightenings.
\begin{figure}[!ht]
\vspace{0.0cm}
\begin{center}
\resizebox{13.0cm}{!}{\plotone{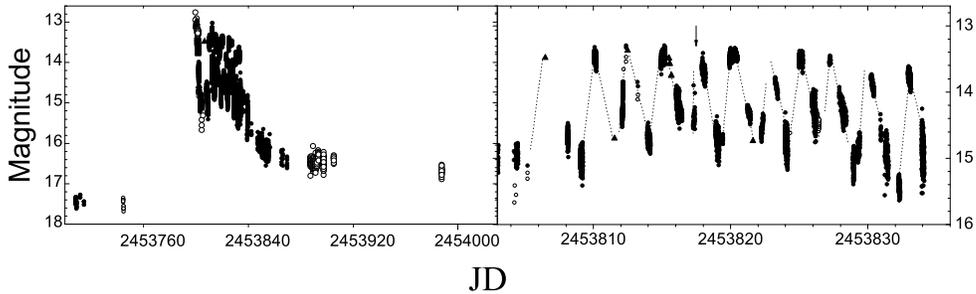}} \vspace{0.0cm}
\caption{Left: The long-term outburst of  the SDSS J0804. {\it R} data are marked by filled circles and {\it V} data - by open circles. Data from AAVSO source (Waagen 2006) are marked by the black triangles. Right: Series of the rebrightenings in detail}
\end{center}
\end{figure}
\section{Prehistory}
The visual inspection of the 900 Sonneberg (1923 -- 2006) and 270 Odessa (1968 -- 1993) archive plates was done. The star was found in the outburst ($\sim 12^{m}.5$) only on one plate of 1979 year from Odessa collection.  Such rare outbursts are typical for the WZ Sge stars.
\section{Spectrum}
The spectrum of the SDSS0804 was obtained during one of the rebrightenings (the date of the observations is marked by arrow in  Fig. 1) in the region 4000 $\AA$ -- 5500 $\AA$. It displays  the  absorbtion lines of Hidrogen. A weak  $H_\beta$ resembles to the P-Cyg profile.
\section{Orbital and superhump periods}
We constructed the periodogram for all data in the outburst after removing trends corresponded for the longer brightness variations, using the Stellingwerf method (Pelt, 1992). It is shown in Fig. 2 (left). The most prominent peak points to the superhump period of 0.059713(7) d. In Fig. 2 (right) the superhump profile evolution during the cource of the outburst is shown. 
\begin{figure}[!ht]
\vspace{0.0cm}
\begin{center}
\resizebox{13cm}{!}{\plottwo{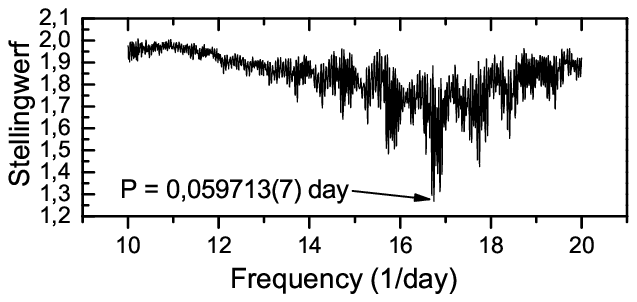}{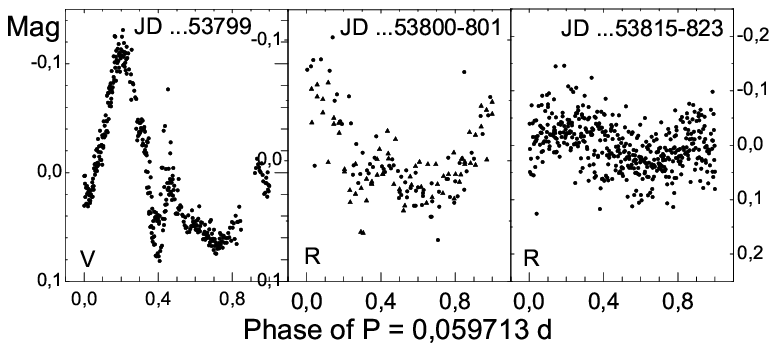}}
\vspace{0.0cm}
\caption{Left: The periodogram for the outburst decline for the {\it R} data pointing to the superhump period. Right: Examples of data folded on the superhump period}
\end{center}
\vspace{0.0cm}
\end{figure}
Prior to rebrightenings the profile has one strong hump and two dip-like structure before and after the hump (see the left and middle panels) at phases 0.02 and 0.40. During rebrightenings the profile changed from cycle to cycle.

The light variations in minimum (V band) on Jan. 8, 2006, correspond to period 0.0586(8) d that fairly well coincides with orbital one. The periodogram and the data convolution are shown in Fig. 3. The mean amplitude
for this night is $0^{m}.1$ and the mean profile is the two-humped one - similar to those obtained by Szkody et al. (2006). Note that the orbital period is also typical to the WZ Sge stars.
\begin{figure}[!ht]
\vspace{0.0cm}
\begin{center}
\resizebox{13.0cm}{!}{\plottwo{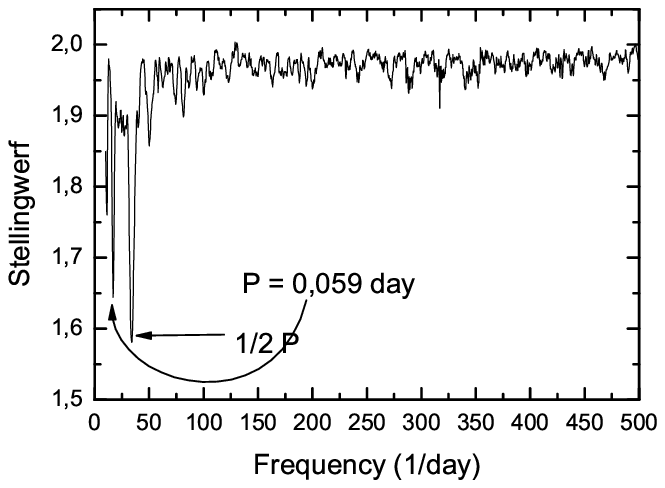}{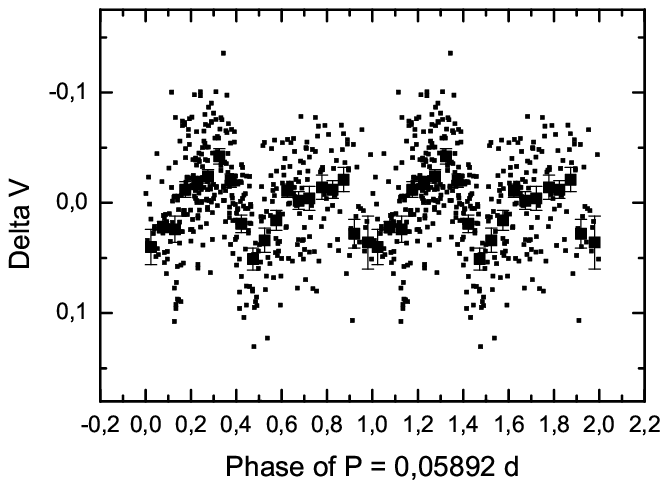}}
\vspace{0.0cm} \caption{The periodogram for the data in the low brightness state in the vicinity of the orbital period (left) and data folded on this period. Mean points obtained by averaging of the data in each of 10 intervale of period are also shown.
(right)}
\end{center}
\vspace{0.0cm}
\end{figure}
For the {\it V} data in maximum (March, 04 and March, 05) obtained with 10-sec time resolution we searched for the short-term light variations. The data for both nights after the superhump profile removed do display the existence of such variations: 7.8 min for 4.03 and $\sim$ 9.0 min for 5.03 with amplitude $\sim 0^{m}.01$ For example the periodogram and data for 4.03 folded on the 7.8 min period are shown in Fig. 4. The source of this QPOs might be the reprocessing by the accretion disk of the hot white dwarf pulsations. 
\begin{figure}[!ht]
\vspace{0.0cm}
\begin{center}
\resizebox{13.0cm}{!}{\plottwo{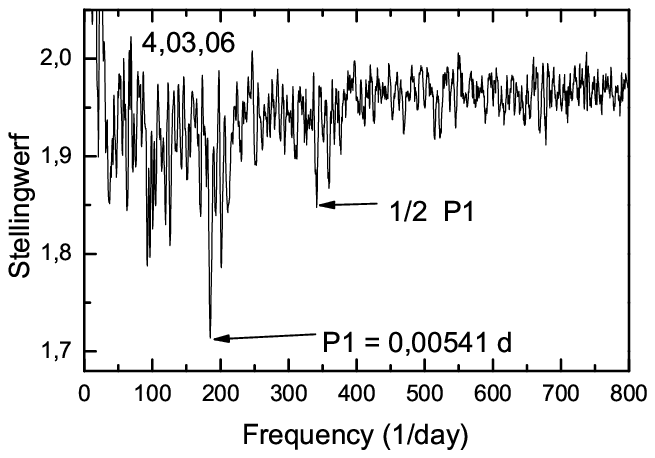}{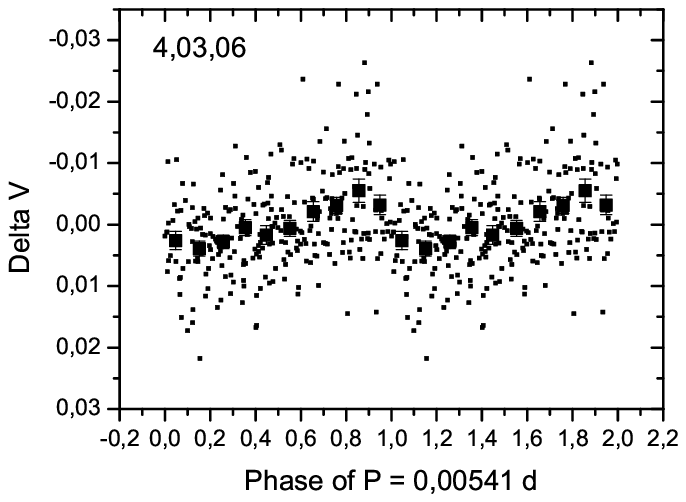}}
\vspace{0.0cm}
\caption{Left: The periodogram for the {\it V} data in the vicinity of
the "`short"' period. Right: Data folded on the 0.00541 d period. The mean points are also shown}
\end{center}
\vspace{0.0cm}
\end{figure}

\section{Acknowledgements} EP, NK, SSh and AG are grateful to the Royal Society for the financial support of the participation in the Workshop. NB is grateful to Afanas'ev for the supplying with the program of spectral data reduction. The work was particularly supported by
grants NSh 5218.2006.2, RFBR 06--02--16411.

\end{document}